\documentclass[aps,prb,twocolumn,sd,amsmath,amssymb,final,superscriptaddress]{revtex4-1}
\usepackage{dcolumn}% Align table columns on decimal point
\usepackage{graphicx}% Include figure files
\usepackage{epstopdf}
\usepackage{amsmath}
\usepackage[bookmarks=false,pdfstartview=FitH]{hyperref}
\usepackage{color}
\usepackage{booktabs}
\usepackage[all]{hypcap}
\usepackage{ragged2e}

\hypersetup{pdftex,
  breaklinks=true,  % so long urls are correctly broken across lines
  colorlinks=true,
  urlcolor=blue,
  linkcolor=blue,
  citecolor=blue,
  }
\def\Let@{\def\\{\notag\math@cr}}
\begin{document}

% Use the \preprint command to place your local institutional report
% number in the upper righthand corner of the title page in preprint mode.
% Multiple \preprint commands are allowed.
% Use the 'preprintnumbers' class option to override journal defaults
% to display numbers if necessary
%\preprint{}

%Title of paper
\title{Accuracy of the HSE hybrid functional to describe many-electron interactions and charge localization in semiconductors}

% repeat the \author .. \affiliation  etc. as needed
% \email, \thanks, \homepage, \altaffiliation all apply to the current
% author. Explanatory text should go in the []'s, actual e-mail
% address or url should go in the {}'s for \email and \homepage.
% Please use the appropriate macro foreach each type of information

% \affiliation command applies to all authors since the last
% \affiliation command. The \affiliation command should follow the
% other information
% \affiliation can be followed by \email, \homepage, \thanks as well.
\author{Mauricio A. Flores}
\email[]{mflores@next-solar.net}
%\homepage[]{Your web page}
%\thanks{}
%\altaffiliation{}
\affiliation{Facultad de Ingenier\'ia y Tecnolog\'ia, Universidad San Sebasti\'an, Bellavista 7, Santiago, 8420524, Chile.}

\author{Walter Orellana}
\affiliation{Departamento de Ciencias F\'isicas, Universidad Andres Bello, Sazi\'e 2212, Santiago, 8370136, Chile.}

\author{Eduardo Men\'endez-Proupin}
%\email[]{Your e-mail address}
%\homepage[]{Your web page}
%\thanks{}
%\altaffiliation{}
\affiliation{Departamento de F\'isica, Facultad de Ciencias, Universidad de Chile, Las Palmeras 3425, \~Nu\~noa, Santiago, 7800003, Chile.}

%Collaboration name if desired (requires use of superscriptaddress
%option in \documentclass). \noaffiliation is required (may also be
%used with the \author command).
%\collaboration can be followed by \email, \homepage, \thanks as well.
%\collaboration{}
%\noaffiliation

\begin{abstract}
Hybrid functionals, which mix a fraction of Hartree-Fock (HF) exchange with local or semilocal exchange, have become increasingly popular in quantum chemistry and computational materials science. Here, we assess the accuracy of the Heyd-Scuseria-Ernzerhof (HSE) hybrid functional to describe many-electron interactions and charge localization in semiconductors. We perform diffusion quantum Monte Carlo (DMC) calculations to obtain the accurate ground-state spin densities of the negatively charged (SiV)$^-$ and the neutral (SiV)$^0$ silicon-vacancy center in diamond, and of the cubic silicon carbide (3C-SiC) with an extra electron. We compare our DMC results with those obtained with the HSE functional and find a good agreement between both methods for (SiV)$^-$ and (SiV)$^0$, whereas the correct description of 3C-SiC with an extra electron crucially depends on the amount of HF exchange included in the functional. Also, we examine the case of the neutral Cd vacancy in CdTe, for which we assess the performance of HSE against the many-body \emph{GW} approximation for the description of the position of the defect states in the band gap.
\end{abstract}

\date{\today}

% insert suggested PACS numbers in braces on next line
%\pacs{}
% insert suggested keywords - APS authors don't need to do this
%\keywords{}

%\maketitle must follow title, authors, abstract, \pacs, and \keywords
\maketitle

\section{Introduction}
Density functional theory (DFT) \cite{Hohenberg64,Kohn99} has become the leading method for electronic structure calculations in materials science, quantum chemistry, and condensed-matter physics.\cite{Hasnip14,Jones15} The DFT formalism shows that ground state properties of a many-electron system can be determined from the knowledge of the electron density distribution alone, thereby avoiding the computation of massively complex many-dimensional wave functions. The ground state density is described by a single determinant with all many-body effects included in one term, the exchange-correlation functional. Unfortunately, the exact functional is unknown and it is necessary to use practical approximations.

Extensive research efforts have been dedicated to obtain better approximations to the exact exchange-correlation functional. Earlier approximations were the local density approximation (LDA), \cite{Kohn65} which considers the exchange and correlation interaction as obtained from a homogeneous electron gas, and the generalized gradient approximation (GGA), \cite{Becke88,Perdew96} which adds gradient terms to the LDA approach. Both approximations are computationally efficient and give reasonable results for ground state properties of molecules and solids, however, they have limitations that hinder their predictive power. They do not properly account for the long-range dispersion forces in van der Waals systems, \cite{Jones79,Kohn98} underestimate the energy barriers of chemical reactions, give erroneous dissociation energies of diatomic radicals, \cite{Mori06} and severely underestimate the band-gap in semiconductors and insulators. \cite{Perdew83} Better approximations can achieved by including additional information to the energy density. In this way, meta-GGAs \cite{Becke98,Tao03} incorporate the second-order gradient of the density, giving accurate results when the system is near mechanical equilibrium, but still failing when bonds are stretched, as nonlocality dominates. \cite{Perdew08} Hybrid-GGAs \cite{Jaramillo03,Heyd03} include nonlocality information by mixing a fraction of Hartree-Fock exchange with local or semilocal exchange. They improve the description of the energy band gap in semiconductors and insulators, but the results critically depend on amount of Hartree-Fock exchange included in the functional. Although some studies have shown that hybrid functionals such as the Heyd-Scuseria-Ernzerhof (HSE) \cite{Heyd03,Heyd06} yield good accuracy for defect energy levels in solids, \cite{Lewis17,Chen17} other calculations revealed serious discrepancies when they are compared with more accurate methods such as diffusion quantum Monte Carlo (DMC). \cite{Santana15,Santana17,Gerosa17} More recently, double-hybrids GGAs have been proposed.\cite{Grimme06,Sharkas11,Su18,Cui18} In addition to a fraction of Hartree-Fock exchange, they include a fraction of second order M{\o}ller-Plesset (MP2) correlation within the random phase approximation (RPA).

Recently, Medvedev and co-workers \cite{Medvedev17} pointed out that DFT is straying from the path toward the exact functional, which should give  ``the right answer for the right reason.''\cite{Korth17,Hammes17} They show that modern highly parameterized functionals have improved the energies while not always improving the electron densities. Here, we investigate the performance of the HSE hybrid functional to describe many-electron interactions and charge localization in semiconductors beyond that of pristine bulk materials. We present a comparison of electron densities calculated using the increasingly popular HSE functional with those obtained with the accurate wave function based DMC method.\cite{Foulkes01,Austin12,Matus16} We assess the accuracy of HSE for the description of defective diamond and cubic silicon carbide (3C-SiC) with an extra electron. Additionally, we compare the performance of HSE with the many-body $GW$ approximation \cite{Hedin65,Hybertsen85} for the description of the position of localized defect levels introduced in defective CdTe. In the latter case, we find that the two approaches qualitatively differ in determining of the position of the anti-bonding orbital introduced by the neutral Cd vacancy with respect to the conduction band edge, irrespective of the amount of Hartree-Fock exchange included in the hybrid functional.

\begin{figure*}
\vspace{0.3cm}
 \centering \includegraphics[width=8.45cm]{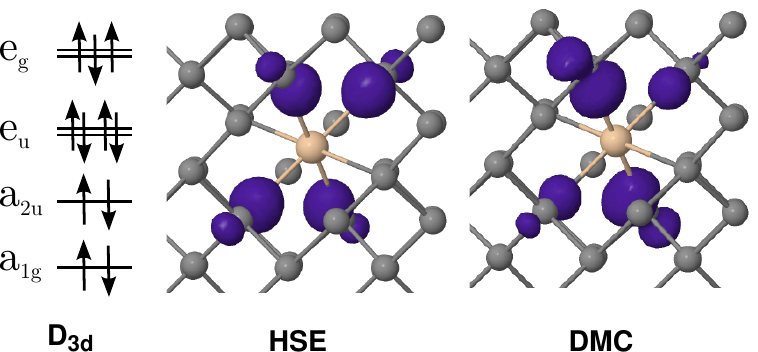} \hspace{0.7cm} \includegraphics[width=8.45cm]{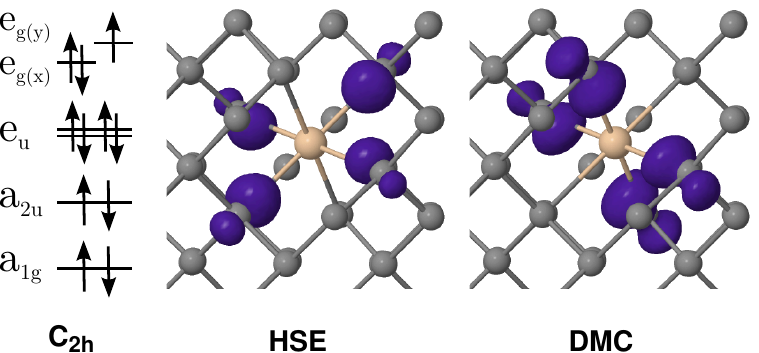}
\caption{Schematic diagram of the electronic structure and the corresponding electron spin density ($\rho$ = 20\% of the maximum) obtained by HSE and DMC calculations of the negatively charged silicon-vacancy center in diamond in D$_{3d}$ (left) and C$_{2h}$ symmetry (right).}\label{fig:1}
\end{figure*}

\section{Methods}

 \subsection{Computational Details}

We performed density functional calculations using the optimized norm-conserving pseudopotential library v0.4 \cite{Van18} generated by Hamann, \cite{Hamann13} in a plane-wave basis set with an energy cutoff of 80 Ry. We used the HSE hybrid functional with the standard exchange and screening parameters, as implemented in the \textsc{\scriptsize{QUANTUM-ESPRESSO}} code. \cite{Giannozzi09,Giannozzi17}

Our DMC calculations were performed using the CASINO code.\cite{Needs09} We used trial wave functions $(\Psi_T)$ of the Slater-Jastrow \cite{Slater29,Jastrow55} type:
\begin{eqnarray}
\Psi_T = \det\{\psi_\uparrow\}\det\{\psi_\downarrow\}e^J,
\label{ec:1}
\end{eqnarray}
where the determinants are composed of single particle orbitals obtained from spin-polarized DFT calculations performed using the \textsc{\scriptsize{QUANTUM-ESPRESSO}} code package.\cite{Giannozzi09,Giannozzi17} We used Trail and Needs norm-conserving pseudopotentials \cite{Trail05,Trail05smooth} with a plane-wave energy cutoff of $\sim 4081.71$ eV (= 150 Ha). Electron terms, electron-nucleus terms, and electron-electron-nucleus terms were included in the Jastrow correlation factor $(e^J)$, whose parameters were optimized through variance minimization of the local energy at the variational Monte Carlo (VMC) level. \cite{Drummond04,Drummond08} This process was followed by DMC calculations within the fixed-node approximation. \cite{Anderson75,Ceperley80,Reynolds82} In this approach, the diffusion process is confined inside the connected nodal region of $\Psi_T$ where the wave function is always positive and vanishes at the boundaries. Unless the nodal surface is exactly correct the fixed-node approximation gives an upper bound to the true ground-state energy. In general, the nodal structure given by PBE or LDA wave functions multiplied by an optimized Jastrow factor to correct for electron correlation are found quite accurate for systems without partially occupied $d$ orbitals. \cite{Leung99,Hood03,Alfe05,Hood12,Kolorenc10}

In our calculations, the trial wave functions were generated from spin-polarized DFT-PBE calculations using 64-atom supercells for diamond and 3C-SiC, at the R-point only. At DMC level, the spin-density was calculated using the mixed estimator. \cite{Foulkes01} We used a DMC time step $\tau$ of 0.01 a.u. and a target population of 5048 walkers, which resulted in an acceptance ratio grater than 99.6\%. To reduce the bias due to pseudopotential localization, we used the T-move scheme proposed by Casula. \cite{Casula06} Additionally, to establish the difference among the methods under study, we calculated the dissociation curve of H$_2$ by placing the molecule in a cubic supercell of side 18 \AA.

For the Cd vacancy in CdTe, we performed many-body $G_0W_0$ calculations using the ABINIT simulation code.\cite{Gonze09,Gonze16} The calculations were performed in 64-atom supercells, using a $2\times2\times2$ \textbf{k}-point mesh to obtain a converged DFT charge density that was then used a as starting point for a subsequent COHSEX (Coulomb-Hole Screened Exchange)+$G_0W_0$ calculation at the $\Gamma$-point only. We employed projector-augmented wave (PAW) \cite{Blochl94} pseudopotentials from the GBRV library \cite{Garrity14} and a plane-wave energy cutoff of 30 Ry. In addition, we used a 20-Ry energy cutoff to represent the dielectric matrix and 3200 bands plus the extrapolar approximation of Bruneval and Gonze. \cite{Bruneval08} The frequency dependence of the dielectric matrix was approximated by the plasmon-pole model of Hybertsen and Louie. \cite{Hybertsen86}

\section{Results and discussion}

\subsection{Many-electron interactions in the silicon-vacancy center in diamond}

Defects in diamond have been proposed as promising candidates to realize single photon sources with applications in quantum cryptography, \cite{Gisin02} magnetic field sensing, \cite{Balasubramanian09,Thiering17} and quantum optics. \cite{Togan10,Aharonovich11,Thiering18_2} In particular, the negatively charged silicon-vacancy center (SiV)$^-$ \cite{Muller14,Becker16,Becker17,Haussler17,Dhomkar18,Ekimov18} has attracted considerable interest due to its exceptional optical properties. The (SiV)$^-$ has a split-vacancy configuration wherein the Si atom adopts a position equidistant from two carbon vacancies. This configuration has D$_{3d}$ symmetry and six carbon dangling bonds, which combine into $a_{1g}$, $a_{2u}$, $e_{u}$, and $e_{g}$ defect orbitals. Eleven electrons fill these orbitals: six electrons are provided by the dangling bonds, four $sp^3$ electrons are donated by the Si impurity and one additional electron is captured by the defect center, leading to the $a^2_{1g}a^2_{2u}e^4_{u}e^3_{g}$ open-shell electronic configuration. As the e$_{g}$ state is partially filled, the (SiV)$^-$ should, in principle, undergo a Jahn-Teller distortion to a lower symmetry point group. However, photoluminescent emission (PL) and excitation (PLE) experiments, \cite{Rogers14} as well as Zeeman studies \cite{Hepp14} suggest that this defect center keeps the D$_{3d}$ symmetry. The origin remains unclear, but an explanation is given in terms of a dynamic Jahn-Teller effect.\cite{Ham65,Gali13,Rogers14,Thiering2018,Londero18}

One way to assess the accuracy of the hybrid functional for the description of many-electron interactions is by computing first the spin-density at HSE level and then comparing the results with those obtained using the more accurate quantum Monte Carlo method. In the latter case, the dependence of the many-electron wave function on electron-electron separations is first taken into account at VMC level by including an optimized Jastrow correlation factor.\cite{Drummond04} Additionally, the projection operation in the subsequent DMC calculation includes all dynamic correlation (whether included or not in the optimized trial function) for states that fall within the space consistent with the fixed node constraint.

\begin{figure}[h!]
 \centering \includegraphics[width=8.5cm]{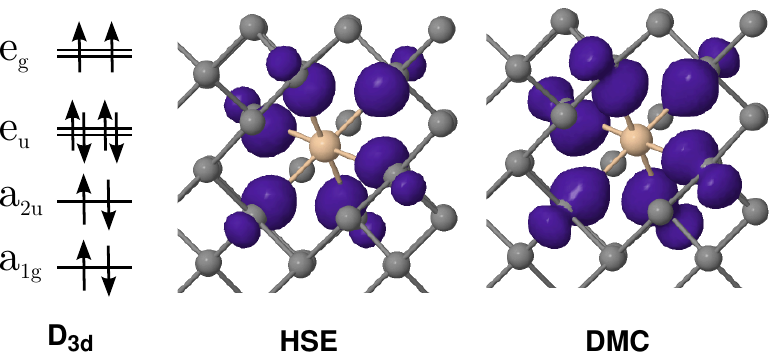}
\caption{Schematic diagram of the electronic structure and the corresponding electron spin density ($\rho$ = 20\% of the maximum) obtained by HSE and DMC calculations of the silicon-vacancy center in diamond in the neutral charge state in D$_{3d}$ symmetry.}\label{fig:2}
\end{figure}

We first investigate the (SiV)$^-$ in the symmetrical D$_{3d}$ configuration, which has orbital and spin degeneracy, and in the C$_{2h}$ configuration where the e$_g$ states are split by a static Jahn-Teller distortion.\cite{Opik57}
In the former case, there are three electrons in the $e_{g}$ level, leaving a single hole with $e$ symmetry and a many-body state with $^2E$ total spin symmetry. The schematic representation of the electronic structure of the D$_{3d}$ configuration, based on a simple molecular orbital model, and  the spin-density obtained by HSE and DMC calculations are shown in Figure \hyperref[fig:1]{1} (left). Both methods give qualitatively the same results. \footnote{The HSE spin-density was obtained by starting the calculation from an asymmetric initial magnetization. A totally symmetric solution (similar to that depicted in Figure 2) can be found if the calculation starts from zero magnetization. In the VASP code, the symmetric solution can be obtained by specifying SIGMA=0.2 (the default value), whereas the asymmetric solution can be found by setting SIGMA=0.02. If the PBE functional is used, the spin-density is totally symmetric, similar to that depicted in Figure \hyperref[fig:2]{2}. In the case of HSE, the inclusion of a fraction of non-local exact exchange breaks the symmetry of the system, lifting the degeneracy of the partially occupied e$_g$ level in the spin-down channel. Moreover, if the system is allowed to relax, it finds its minimum in C$_{2h}$ symmetry. In DMC, the inclusion of the spatially dependent correlation energy (i.e., electron-electron interactions beyond their mean-field average) breaks the spin degeneracy of the single-particle e$_g$ states. As the resulting spin-density is inconsistent with the symmetry mandated by the D$_{3d}$ point group, a structural distortion to a lower symmetry group is suggested.}

 According to the Jahn-Teller theorem, \cite{Jahn37} the orbital degeneracy of the partially occupied $e_{g}$ level can be lifted by a small structural distortion to a less symmetric point group. We found two configurations with  C$_{2h}$ symmetry where two carbon dangling bonds are located ${\sim 0.03}$ \AA \hspace{0.01cm} closer and ${\sim 0.03}$ \AA \hspace{0.01cm} away from the silicon impurity. At DMC level,  the former configuration is more stable by about 0.12 eV. Figure \hyperref[fig:1]{1} (right) shows the schematic representation of the electronic structure of this configuration and the corresponding spin-density obtained by HSE and DMC calculations. The HSE result shows that the spin-density is localized equally on the four C atoms located farther away from the Si atom. On the other hand, DMC indicates that spin-density is localized on the two C atoms located closer to the Si atom and on two of the four C atoms located farther away from the impurity. \cite{mflores_files} Nevertheless, the HSE functional can correctly describe the localization of a hole on the $e_{g_(y)}$ state. It is worth mentioning the recent work of Bockstedte et al., \cite{Bockstedte18} that highlights the importance of dynamic correlation effects on the energy-level structure of the excited state of the nitrogen-vacancy center in diamond.

It is also suitable to investigate the silicon-vacancy in the neutral charge state (SiV)$^0$. This complex has a $a^2_{1g}a^2_{2u}e^4_{u}e^2_{g}$ electronic configuration which according to the Hund's rule results in a S=1 ground-state electron spin configuration. \cite{Green17} The schematic representation of the electronic structure as well as a comparison between the spin-density obtained by HSE and DMC calculations are shown in Figure \hyperref[fig:2]{2}. In contrast to (SiV)$^-$, we observe a qualitatively agreement between HSE and DMC. Although the HSE spin-density is slightly over localized as compared with the DMC result, in both cases the distribution has its maxima equally centered on the six carbon dangling bonds.

Next, we discuss about how many-electron interactions are treated in the Kohn-Sham density functional theory and within the DMC approximation. In the case of electrons with parallel spins, the antisymmetry requirement of the wave function fulfills the Pauli exclusion principle and keeps electrons apart introducing the so-called Fermi hole. \cite{Giner2016} However, the Pauli exclusion principle exerts only a small influence in electrons of opposite spin and thus Coulomb interactions should be explicitly taken into account. Therefore, we should note that correlation errors mainly affect the description of electrons having opposite spins. In DFT, electron interactions are approximated by a mean-field potential and the many-body wave function is represented by a single Slater determinant composed of single-particle orbitals. This is a good approximation for systems of slowly varying density, but it is insufficient for situations of degeneracy or near degeneracy, \cite{Becke03,Cohen08,Cohen08_2} or when electron correlations depend on the shape of the ground-state vacant orbitals or excited-state orbitals. In these cases, the system is poorly described by a single Slater determinant. On the other hand, in DMC the effect of short-range correlations is accounted by a Jastrow correlation factor, which reduces the amplitude of the wave function when electrons are close to each other, thereby making the wave function explicitly dependent on the position.

\begin{figure}[h!]
\vspace{0.3cm}
 \centering \includegraphics[width=7cm]{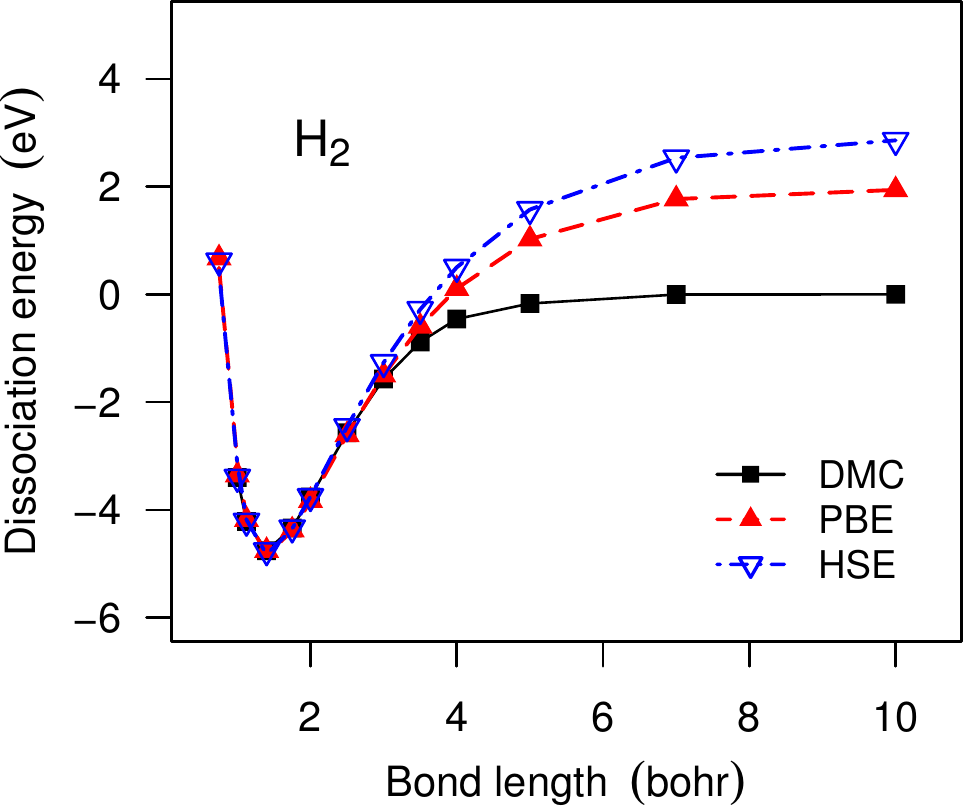}
\caption{Dissociation energy curve of the H$_2$ molecule obtained by DMC, PBE, and HSE calculations.}\label{fig:3}
\end{figure}

The difference between the two approximations can be clearly seen by considering the stretched H$_2$ molecule, the simplest case of a strongly correlated system, where the true (interacting) ground state is a singlet. \cite{Kolos60} The ground state energy of the hydrogen molecule was previously calculated by Traynor and Anderson \cite{Traynor91} using the quantum Monte Carlo method. Our results for the dissociation curve of the H$_2$ molecule obtained within the DFT approximation (employing both PBE and HSE exchange-correlation functionals) and with the DMC method are shown in Figure \hyperref[fig:3]{3}. According to the mean-field approximation of DFT, the probability of finding both electrons in the same atom is always $1/2$, regardless of the bond length. \cite{Olsen14} This is a good approximation when the atoms are near their equilibrium bond length ($\sim$1.4 bohr), but it fails in the dissociation limit where each H atom has a half spin-up electron and a half spin-down electron. \cite{Cohen08_2}

The reason behind the good accuracy of HSE in the description of (SiV)$^0$ might be explained by the fact that in this case there are two electrons with parallel spins occupying the $e_{g}$ orbital. Moreover, it was demonstrated that this many-body state can always be represented by a single Slater determinant. \cite{Lowdin59} In contrast, the (SiV)$^-$ center has two electrons of opposite spins occupying the $e_{g}$ orbital, thus electron-electron interactions beyond their mean-field average may become important.

\subsection{Small polarons in 3C-SiC}

\begin{figure}[h!]%
\centering \includegraphics[height=5cm]{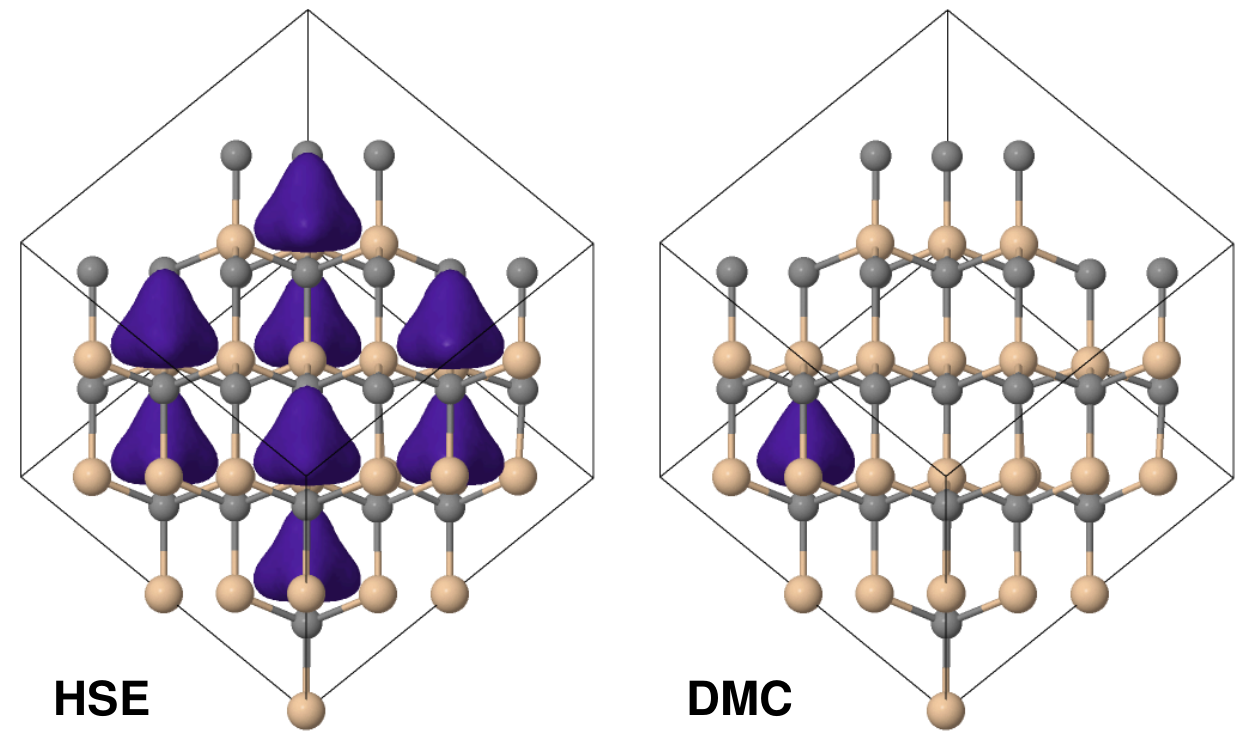}
\caption{Electron spin density ($\rho$ = 40\%  of the maximum) of 3C-SIC obtained by HSE (left) and DMC (right) calculations. Beige and turquoise spheres represent Si and C atoms, respectively.}
\end{figure}\label{fig:4}

\begin{figure*}%
 \centering
 \includegraphics[width=3.8cm]{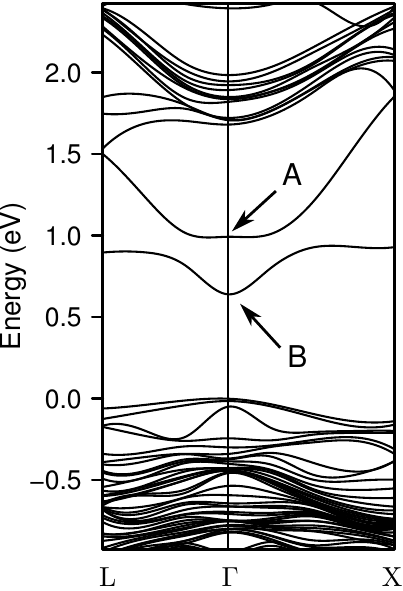}\hspace{-0.1cm}
  \includegraphics[width=3.9cm]{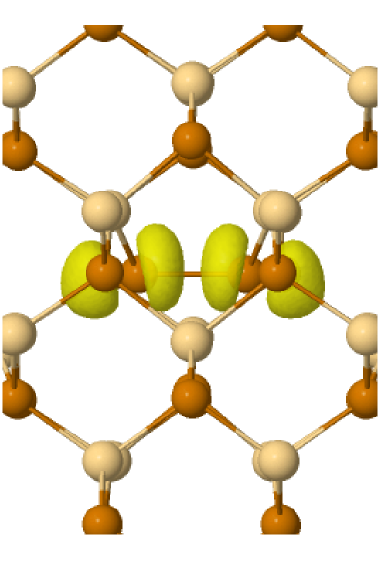}\hspace{0.8cm}
  \includegraphics[width=3.8cm]{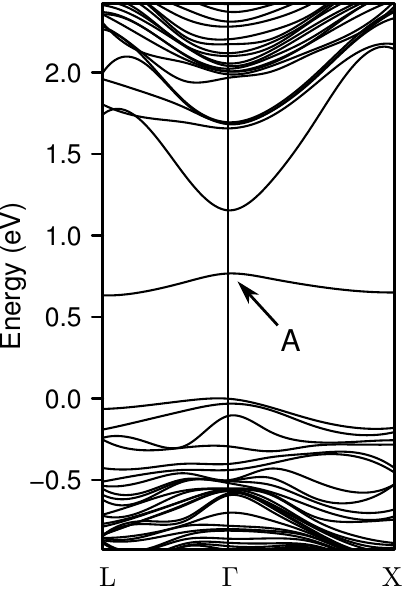}\hspace{-0.1cm}
  \includegraphics[width=3.9cm]{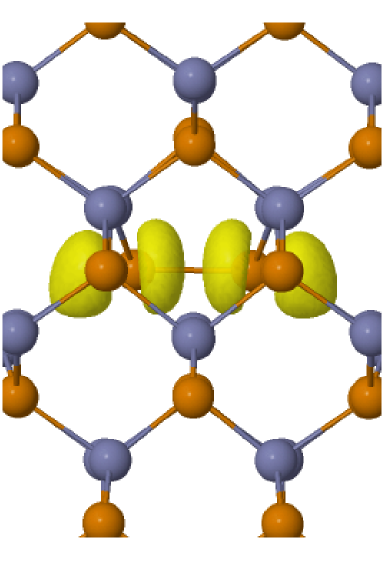}
 \caption{Electronic band structures and charge density isosurfaces ($\rho = 0.002e/\text{bohr}^3$) corresponding to the energy level labeled as A of the neutral cation vacancy in CdTe (left) and ZnTe (right), in C$_{2v}$ symmetry. Brown spheres represent Te atoms; beige and azure spheres represent Cd and Zn atoms, respectively. The calculations were performed at DFT-PBE level in a 128-atom supercell.}\label{fig:5}
\end{figure*}

It was recently pointed out that the inclusion of a fraction of Hartree-Fock interaction in hybrid functionals may introduce a spurious exchange splitting between occupied and unoccupied states. \cite{Jang12,Bang13,Gao16} In spin-polarized calculations, this effect could lead to unusually large magnetic moments,\cite{Jang12} wrong magnetic ground-states, \cite{Grau12} and a slow convergence of the total energy of the system with respect to the supercell size. \cite{Broqvist09,Flores17}

In the following, we investigate the effects of the inclusion of Hartree-Fock interaction in the HSE functional by considering the negatively charged 3C-SiC. The excess electron may: (1) self-localize by coupling to a lattice distortion, forming a polaron; (b) retain its free-carrier character. For TiO$_2$, it was found that the solution crucially depends on the fraction of the Hartree-Fock exchange included in the exchange-correlation functional.\cite{Janotti13,Elmaslmane18}

We performed spin-polarized calculations to determine the ground state of the negatively charged 3C-SiC system. We added an extra electron to the conduction band and then allowed the system to relax. Our calculations were performed in a 64-atom supercell, using the HSE functional ($\alpha = 25\%$) for structural relaxations. We found a slightly distorted structure in which one C-Si bond is stretched by $\sim$0.07 \AA.\footnote{The distorted 3C-SiC structure is included as  Supplementary Material.}  In Figure \hyperref[fig:4]{4}, we compare the spin-density obtained by HSE and DMC calculations. We observe that both approximations give qualitatively different results. According to HSE, the negatively charged 3C-SiC is an electride-like compound \cite{Dye09,Dong17,Flores18_1} which is characterized by charge localization at interstitial regions; on the other hand, DMC predicts the formation of a small polaron. The HSE wave function has large magnetic moments on the Si atoms near the local maxima of the spin-density and exhibits negative values on the nearby C atoms. We found that the unusual HSE result is due to a finite size effect, a direct consequence of the long-range nature of the Hartree-Fock exchange that converges extremely slowly with respect to the supercell size. The same result as shown in Figure \hyperref[fig:4]{4} is obtained using a 2$\times$2$\times$2 k-point mesh displaced from the $\Gamma$-point, which is equivalent to a single k-point calculation in a 512-atom supercell. However, if we decrease the fraction of the Hartree-Fock exchange to 10\% or use a finer \textbf{k}-point mesh, we find that the spin-density is in close agreement with the DMC result. The form of the long-range tail of the Coulomb interaction critically depends on the screening parameter ($\mu$) which defines the extent of the Hartree-Fock exchange in real space \cite{Skone14} Therefore, we strongly recommend the use of large simulation cells and a careful choice of the amount of Hartree-Fock exchange included in the hybrid functional.

\subsection{Localized defect levels in the band gap}

In the simulation of point defects in semiconductors and insulators, a correct description of the band gap as well as the absolute position of the band edges is of critical importance. \cite{Du2015,Flores16} The severe underestimation of the band gap in standard DFT has lead to large discrepancies and conflicting results among theoretical calculations. Hybrid functionals can give reliable band gaps for most semiconductors when an optimum system-dependent fraction of Hartree-Fock exchange energy is used.\cite{Heyd05} Moreover, it is commonly assumed that they can give a reliable description of defects states,\cite{Heyd05,Chen17,Gerosa17} although the use of the same parameters to describe orbitals with a distinct degrees of localization seems questionable. In the following, we assess the accuracy of the HSE functional in the not uncommon case when a localized defect state merges with one of the band edges of the host. We consider the Cd vacancy in CdTe, which is among the most important native defects in this semiconductor.

The Cd vacancy in CdTe has been extensively investigated both theoretically and experimentally; \cite{Emanuelsson93,Shepidchenko15} still several aspects remain unclear. In the neutral charge state, the Cd vacancy undergoes a structural distortion from T$_d$ to C$_{2v}$ symmetry. The situation is similar to the well-known AX distortion,  \cite{Biswas11,Flores17} where two Te atoms move toward each other to form a new bond. The net result is the loss of one bond (as each Te atom breaks one bond with a Cd atom) and the creation of an empty anti-bonding orbital. At DFT-PBE level, this anti-bonding state is strongly hybridized with the conduction-band minimum (CBM), as shown in Figure \hyperref[fig:5]{5} (left). However, this effect is likely to be an artifact of DFT due the fact that the CdTe band gap is too small such that the anti-bonding level erroneously lies above the CBM at the $\Gamma$-point. This can be seen by considering the isoelectronic case of the neutral Zn vacancy in ZnTe where the anti-bonding state appears well isolated in the band gap.

In Figure \hyperref[fig:5]{5}, we compare the DFT-PBE band structure and the squared wave function corresponding to the anti-bonding level (labeled as A) of the neutral cation vacancy in CdTe and ZnTe. In contrast to CdTe, in the case of ZnTe the unoccupied anti-bonding level lies isolated in the band gap of the host, as the CBM is higher in energy. To accurately obtain the position of this anti-bonding level introduced by the Cd vacancy in CdTe, we performed many-body $GW$ calculations that can give accurate band structures of solids. \cite{Flores17_1,Flores17_2,Klimevs14} We used DFT-PBE wave functions as a starting point for a subsequent COHSEX$+G_0W_0$ perturbative calculation. We found that the unoccupied anti-bonding state lies 0.26 eV below the CBM.

Taking the $G_0W_0$ result as a reference, a natural question arises: is the HSE hybrid functional able to correctly describe the position of the anti-bonding orbital associated with the neutral Cd vacancy in CdTe? We performed HSE calculations using 128- and 250-atom supercells in which the Brillouin zone was sampled at the $\Gamma-$point only. These calculations were performed using PAW potentials, as implemented in the Vienna \emph{ab initio} simulation package
(VASP).\cite{Kresse96} We found that HSE gives a band gap of 1.41 eV, in good agrement with the COHSEX+$G_0W_0$ result of 1.42 eV. However, HSE keeps the PBE result by placing the anti-bonding orbital 0.48 eV above the CBM. A similar behavior was previously reported for ZnO:Co, \cite{Walsh08,Su2017} where DFT results are incorrect for the optical absorption and the Co $d$-$d$ splitting is overestimated in the order of 300\% when the fraction of Hartree-Fock exchange corresponds to the one necessary to reproduce the experimental band gap of ZnO.\cite{Walsh08} Another example is the carbon vacancy in 3C-SiC, where the position of the localized defect state erroneously appears above the CBM. \cite{Bruneval12}

The failure of the HSE to describe the position of the anti-bonding orbital in the neutral Cd vacancy in CdTe is an example of the drawback of the uniform treatment of electronic states with distinct degrees of localization. This issue may be corrected through the addition of empirical nonlocal external potentials, \cite{Lany08} by using local hybrid functionals that use a position-dependent mixture of local and exact exchange,\cite{Krukau08,Arbuznikov09} or the orbital dependent exact exchange extension of hybrid functionals proposed by Iv\'{a}dy \emph{et al.}  \cite{Ivady14,Ivady17} The latter approach has the advantage that the orbital dependent parameter can be obtained self-consistently based on the analogy of quasi-particle equations and hybrid-DFT single particle equations. \cite{Ivady17} A recommendation to practitioners studying
defects in semiconductors or any system having states with different degrees of localization is always to plot the band structure of the system (at least using computationally efficient local or semilocal functionals) to identify problems analogous to the presented in this section. In these cases the use of more advanced methods such as $G_0W_0$ is highly recommended.

\section{Summary}

We have investigated the accuracy of the HSE hybrid exchange-correlation functional in the description of three systems with different levels of complexity: the negatively charged and the neutral silicon-vacancy center in diamond, pristine 3C-SiC with an extra electron, and the neutral Cd vacancy in CdTe.

Our results show that the HSE functional can accurately describe many-electron interactions in systems with moderate correlations, such as the neutral and the negatively charged silicon-vacancy center in diamond.  Moreover, the application of the HSE functional to systems with different degrees of localization shows systematic errors: (1) Due to the slow convergence of the Hartree-Fock exchange with respect to the supercell size, it may predict an incorrect ground state for the negatively charged 3C-SiC; (2) When localized defect states artificially merge with delocalized (Bloch-like) conduction-band states at DFT-PBE level, as in the case of the neutral Cd vacancy in CdTe, HSE corrects the band gap but keeps the spurious hybridization given by PBE, predicting an incorrect position of the localized level with respect to the CBM.

\begin{acknowledgments}
This paper was supported by the Fondo Nacional de Investigaciones Cient\'ificas y Tecnol\'ogicas (FONDECYT, Chile)
under Grants No. 1170480 (W.O.) and No. 1171807 (M.A.F. and E.M-P.). Powered@NLHPC: This research was partially supported by the supercomputing infrastructure of the NLHPC (ECM-02).
\end{acknowledgments}

\bibliographystyle{apsrev4-1}
\bibliography{bib}

\end{document}